# Phonons in doped and undoped BaFe$_2$As$_2$ investigated by inelastic x-ray scattering

D. Reznik[1], K. Lokshin[2], D. C. Mitchell[3], D. Parshall[3], W. Dmowski[2], D. Lamago[1,4], R. Heid[1], K.-P. Bohnen[1], A.S. Sefat[5], M. A. McGuire[5], B. C. Sales[5], D. G. Mandrus[5], A. Subedi[3,5], D. J. Singh[5], A. Alatas[6], M. H. Upton[6], A. H. Said[6], A. Cunsolo[6], Yu. Shvyd'ko[6], and T. Egami[2,3,5]

[1] *Forschungszentrum Karlsruhe, Institut für Festkörperphysik, P.O.B. 3640, D-76021 Karlsruhe, Germany*
[2] *Departement of Materials Science and Engineering, University of Tennessee, Knoxville, TN 37996, USA*
[3] *Department of Physics and Astronomy, University of Tennessee, Knoxville, TN 37996, USA*
[4] *Laboratoire Léon Brillouin, CEA Saclay, F-91191 Gif-sur-Yvette, France*
[5] *Oak Ridge National Laboratory, Oak Ridge, TN 37831, USA*
[6] *Advanced Photon Source, Argonne National Laboratory, Argonne, IL 60192, USA*

We measured phonon frequencies and linewidths in doped and undoped BaFe$_2$As$_2$ single crystals by inelastic x-ray scattering and compared our results with density functional theory (DFT) calculations. In agreement with previous work, the calculated frequencies of some phonons depended on whether the ground state was magnetic or not and, in the former case, whether phonon wavevector was parallel or perpendicular to the magnetic ordering wavevector. The experimental results agreed better with the magnetic calculation than with zero Fe moment calculations, except the peak splitting expected due to magnetic domain twinning was not observed. Furthermore, phonon frequencies were unaffected by the breakdown of the magnetic ground state due to either doping or increased temperature. Based on these results we propose that phonons strongly couple not to the static order, but to high frequency magnetic fluctuations.

The recent discovery of superconductivity in iron pnictides [1] with the critical temperature, $T_C$, up to 55 K [2] created huge excitement, since the superconductivity in these compounds is clearly unconventional, and, just as in the cuprates, cannot be explained by the standard BCS theory. They are poor metals with low density of states at the Fermi energy [3], and magnetic and structural phase transitions appear in compositions in close proximity [4-7]. After the initial discovery [1] superconductivity was observed in a large number of closely related compounds. They all share the FeAs layers, in which Fe is surrounded by four As ions forming a tetrahedron, and FeAs$_4$ tetrahedra are connected by sharing the faces. The FeAs layers are separated by ionic layers, such as REO$_{1-x}$F$_x$ (RE = rare earths) or Ba$_{1-y}$K$_y$. Density functional theory (DFT) calculations obtain a very weak conventional electron-phonon coupling [8], which is not sufficient to explain the observed high values of $T_C$.

The DFT calculations as well as experiments show remarkable sensitivity of superconductivity and magnetism to the lattice. Changing the ionic size of the rare-earth without changing the apparent charge density, changes $T_C$ by a factor of two [2], and the appearance of magnetism is closely tied to a very small lattice deformation. [3,9] In addition, there is a large overestimate of the magnetic moments in generalized gradient approximation (GGA) calculations, which appears essential for obtaining a correct atomic position of As in structure optimizations [10,11]. This behavior persists into the non-magnetic superconducting phase. Furthermore, the results of core level spectroscopy show exchange splitting even in the non-magnetic phases [12], similar to the Fe$_2$Nb, which is near an itinerant magnetic quantum critical point [13].

Somewhat surprisingly, certain phonon frequencies calculated using DFT depend very sensitively on the ordered moment. Nonmagnetic calculations where the magnetic moment is fixed to zero produce a poor agreement with experiment for phonons around 20meV. T. Yildirim improved the agreement between calculated and experimental phonon density of states of BaFe$_2$As$_2$ powder by including magnetic order in the calculation. [18] However, the ordered magnetic moment that comes out of such calculations is much larger than the experimentally observed moment even in undoped antiferromagnetic samples. It is zero in most doped supercondcucting compounds. In order to resolve this apparent contradiction, it has been proposed that the magnetic state is frustrated or dynamic [11,17]. Ref. 17 proposed that quasistatic ordered moments are always present in the pnictides, but cannot be easily observed due to the moving domain boundaries. Such dynamic behavior should appear as a quasielastic peak in neutron scattering spectra if the timescale is relatively slow (a few meV), whereas such a peak has not been observed. For faster motion such a peak would blend into the background and become difficult to observe. In this letter we show that in DFT calculations of single crystal phonon spectra frequencies of some modes around 23meV depend not only on the ordered magnetic moment but also on the orientation of the phonon wavevector relative to the magnetic ordering wavevector. Twinning of magnetic domains would result in splitting of these phonon peaks that is proportional to the size of the ordered moments.

We measured phonon frequencies and linewidths by inelastic x-ray scattering (IXS) in double-layered FeAs compounds, BaFe$_2$As$_2$, Ba$_{1-x}$K$_x$Fe$_2$As$_2$ ($x$ = 0.2, 0.4), and BaFe$_{1.8}$Co$_{0.2}$As$_2$, and compared the results to magnetic and nonmagnetic DFT calculations. We found a much better overall agreement of the experimental results with the magnetic calculation, but there was no peak splitting, which was predicted to be substantial for some measured phonons. The magnetic moment that came out of the calculation was also much larger than observed in experiments [4] in agreement with previous work. [8-11]

The IXS measurements were carried out at the Advanced Photon Source (APS), Argonne National Laboratory. The first main insights were obtained based on the measurements on $Ba_{0.8}K_{0.2}Fe_2As_2$ at the XOR 30-ID (HERIX) beamline and were confirmed by more detailed measurements on the other three samples, $BaFe_2As_2$ (FeAs), $Ba_{0.6}K_{0.4}Fe_2As_2$ (BaK), and $BaFe_{1.84}Co_{0.16}As_2$ (FeCo), on the XOR 3-ID beamline. A third set of experiments in both transmission and reflection geometries covered most of the reciprocal space for another $BaFe_{1.84}Co_{0.16}As_2$ crystal and a smaller range of wavevectors for a $BaFe_2As_2$ crystal. The incident energy was about 23.8 keV (30-ID) or 21.66 (3-ID), and the horizontally scattered beam was analyzed by a horizontal array of nine/four spherically curved silicon crystal analyzers on 30-ID/3-ID, respectively. The energy resolution was 1.5/2.2 meV full width at half maximum (FWHM) on 30-ID/3-ID, respectively. The focused beam size was about 30/100 $\mu m$ on 30-ID/3-ID. The FeAs crystals were nonsuperconducting, whereas the doped samples had $T_c$s of 22K (FeCo) and 30 K (BaK). In the preparation of crystals, high purity elements (> 99.9 %) were used and the source of all elements was Alfa Aesar. BaK crystals were grown out of a Sn flux. The procedure involved placing elements in the ratio of $[Ba_{0.6}K_{0.8}Fe_2As_2]$:Sn = 5:95, in an alumina crucible, and sealing into a fused silica tube under partial argon atmosphere (~1/3 atm). The sealed ampoule was heated in a furnace at 800 C for 6 hours, then cooled (10 °C/hr) to 525 °C at which point the Sn was decanted off by means of a centrifuge. Single crystals of FeAs and FeCo were grown out of FeAs flux [14]. For FeAs, crystals prepared with a ratio of Ba:FeAs = 1:5 were heated at 1180°C as described above, for 8 hours. For FeCo, crystals with a ratio of Ba:FeAs:CoAs = 1:4.45:0.55 were similarly heated to 1180 °C, and dwelled for 10 hours. Both ampoules were cooled at the rate of 3 °C/hour, followed by decanting of FeAs flux at 1090 °C. All crystals were well formed plates with smooth surfaces with typical dimensions of ~ 2 x 3 x 0.1 $mm^3$ and the [001] direction perpendicular to the plane of the crystals. Measurements were performed in reflection and transmission modes. Closed cycle He refrigerator was used for temperature control.

The calculations were carried out using the generalized gradient approximation (GGA) and the linear-response technique. They give the phonon frequencies and their eigenvectors, which determine experimentally observed phonon intensities. Details of the calculations are described elsewhere. [3,10,15] We first calculated the phonon spectra for a nonmagnetic ground state and with an optimized structure. In the second calculation both the Fe antiferromagnetic moment and the structural parameters of the doubled unit cell were optimized. The third calculation fixed the Fe moment to zero and the structure to the one obtained in the second calculation. It allowed us to separate the effect of magnetic moment on the phonons from the effect of different structures. The twinning of orthorhombic domains in the calculations for the orthorhombic structure was assumed for the comparison with the experiment, i.e. both orthorhombic directions were superimposed. As a result of the unit cell doubling and twinning the number of branches in the original nonmagnetic tetragonal unit cell is quadrupled compared with the first nonmagnetic calculation.

Figure 1(a,b) shows the experimentally measured phonon frequencies throughout the Brillouin zone (BZ) superimposed with calculations. The first nonmagnetic calculation agrees well with experiment, except the calculated frequencies around 23 meV are 20% too high. Aside from increasing the number of branches due to unit cell doubling, the main effect of magnetism is to soften calculated frequencies around 23 meV and bring them close to the experimental values. Figure 1c compares the magnetic calculation with the nonmagnetic one done for the same structure. Based on inspection of figure 1, the magnetic calculation has a better overall agreement with experiment than both nonmagnetic calculations.

Figure 2 compares the calculated and experimental phonon spectra for the first (nonmagnetic) and the second (magnetic) calculations. Both twin domains were included in the calculated curves. Peak intensitites were obtained from the calculated structure factors, and linewidths were broadened by experimental resolution. The agreement between the nonmagnetic calculation and the data is very good for phonons around 9-15meV and 30-35meV, but the phonons around 23meV are softer than calculated. Magnetism dramatically improves agreement with the experiment, but not completely: The calculation predicts split peaks of the 25 meV phonons in the [110] direction (at $Q=(2.5,1.5,0)$ and $Q=(2.5,2.5,4)$), whereas a single nearly resolution-limited sharp peak is observed. (Fig. 2) This splitting occurs because calculated frequencies of phonons with wavevectors along [110] depend on the orientation of the phonon wavevector relative to the Fe moment.

Figure 3a illustrates that for the [110] direction, the phonon wavevector may be either parallel or perpendicular to the Fe moments depending on the domain, whereas for the [100] or [001] directions the angle between the wavevector and the Fe moments is always the same. Figure 3(b-d) zooms in on the phonons that should be most sensitive to the magnetic order and includes predictions of all three calculations. The magnetic calculation agrees best with the data aside from the unobserved peak splitting as discussed above. The orthorhombic nonmagnetic (third) calculation (green lines in Fig. 3) improves the agreement somewhat over the nonmagnetic calculation with the optimized tetragonal structure, but is clearly inadequate to explain the data. For example, for Q=(2.5,1.5,0), it pulls the 28meV phonon (red) down to the experimental value of 25.5 meV, but at the same time, upper phonon is pulled down from the experimental value of 32meV to 30 meV. Peak splitting in this calculation due to unequal bond lengths in the [110] and [1-10] directions (in the tetragonal notation) is much smaller than the experimental resolution and is not apparent in the simulated spectra. Thus large Fe moment is essential to make the calculated phonon frequencies agree with experiment, the important caveat being that the accompanying peak splitting is not observed. We would like to note here that in the case of moving domain walls, the magnetic order should still split these phonons if the timescale of the motion were slower than the phonon frequency, which is around 20meV.

The destruction of the magnetic order by either doping or temperature has almost no influence on the phonons. (Figure 4 shows a possible small hardening of at most 1meV of some high energy modes upon doping.) We also found that superconductivity appears to only weakly, if at all, influence the phonons that we had measured (data not shown).

Our results seem to be contradictory. On the one hand, the substantial improvement of the agreement of the calculated phonon frequencies with experiment upon inclusion of magnetism speaks in favor of large antiferromagnetic Fe moments. On the other hand, absence of split peaks points to the opposite conclusion. It is also the fundamental feature of the GGA [19] that the calculated optimal length of the Fe-As bond is sensitive to the angle it makes with the Fe moment. It is this bond that is modulated by the phonons around 25meV. Thus it is not possible to somehow "tune" the calculation to remove the magnetism-induced splitting but preserve the softening in the [110] direction. Here we would like to propose a way out of this paradox.

In the doped compounds without magnetic order, the average moment is zero. However, the comparison of measured and calculated structural and vibrational properties indicate the bonding is best described in magnetic calculations with a substantial moment, which can be understood in terms of a large renomalization of the moment by spin fluctuations. In a case where the fluctuations are fast compared with the lattice response (phonon frequencies), the measured static moment is $<m>$, which is zero for the doped case, but for the bonding the relavent quantity is the root mean square: $<m^2>^{1/2}$. These two quantities are of course practically the same in most materials, but in the pnictides with their large amplitude spin fluctuations they are apparently quite different according to the above arguments. In magnetic DFT calculations, the moments are necessarily ordered along some direction, leading to peak splitting. However, by symmetry, in the doped compounds, the rms moment will be isotropic globally. Moving quasistatic domains would break symmetry locally while keeping $<m>$ at zero. But in this case the phonon peaks would be split as discussed above. The absence of peak splitting in experiment is inconsistent with the idea of quasistatic domains, unless the moments fluctuate faster than phonon frequencies. Absense of peak splitting (or weak splitting) in the undoped magnetic phase, where the symmetry argument does not apply indicates that these fast fluctuations rather than static ordering also dominate this phase. The main contribution to rms moment is from fluctuations rather than the ordered moment, which is considerably reduced (i.e. renormalized) as compared to the static DFT value. Consistent with this idea, recent neutron scattering experiments have shown that high frequency fluctuations concentrated primarily around the antiferromagnetic wavevector and extending up to 200meV dominate the magnetic spectrum of the pnictides at all investigated doping levels. [20] Since the fluctuations are much faster than the phonon frequencies, the phonons would sense the effect of antiferromagnetism averaged between the two AF wavevetors ([0.5,0.5,0] and [0.5,-0.5,0] in tetragonal notation) and would not be split. Based on these observations, we conjecture that phonons soften due to a coupling to these high frequency fluctuations rather than to static or quasistatic moments. This mechanism does not seem to be captured in the DFT and its theoretical underpinning is still to be developed.

Although the above analysis involved only the $BaFe_2As_2$-based family of pnictides, we expect it to apply to all members of the pnictide family. In $CaFe_2As_2$ magnetic calculations also predicted splitting of some phonons, which was not observed. [15] In contrast with $BaFe_2As_2$, magnetic calculations for $CaFe_2As_2$ did not improve the agreement with experiment compared with nonmagnetic calculations for the same structure. The main difference between $BaFe_2As_2$ and $CaFe_2As_2$, is that the latter is close to the so-called collapsed phase (which can be achieved by applying only a few kbar), whereas the $BaFe_2As_2$ and other pnictides are not. Ref. [21] seems to conclude that the magnetic calculation does give a better agreement for $CaFe_2As_2$. However, this improvement with respect to the non-magnetic case could be due to the different values for the optimized internal structural parameter, which is much closer to the experimental value in the case of the magnetic calculation. According to ref. [15] the improvement obtained for the phonons in this case comes entirely from the difference in the optimized structures, not due to magnetism.

Fukuda, *et al*. [16] measured phonons in single-layer compounds $LaFeAsO_{1-x}F_x$ ($x = 0$, 0.1) and $PrFeAsO_{1-y}$ ($y \sim 0.1$), but not along the [1 1 0] direction where peak splitting is expected. Decreasing the Fe-As coupling strength by 10% compared to a nonmagnetic LDA calculation improved the agreement with experiment, but recently they found that the magnetic calculation works even better than the force constant modification. [16]

To summarise, the detailed comparison of experimental and calculated phonon dispersions confirms the overall idea that phonon frequencies are influenced by Fe moments. However, the scenario of moving magnetic domain walls does not explain the absence of phonon peak splittings in the [110] direction. We propose an alternative picture of phonons coupling to high frequency magnetic fluctuations.

This research was supported in part by the Department of Energy EPSCoR Implementation award, DE-FG02-08ER46528 to the University of Tennessee, and by the Basic Energy Science Division of the Department of Energy. Use of the Advanced Photon Source was supported by the U. S. Department of Energy, Office of Science, Office of Basic Energy Sciences, under Contract No. DE-AC02-06CH11357. Work at ORNL was supported by DOE, Division of Materials Sciences and Engineering The construction of HERIX was partially supported by the NSF under grant no. DMR-0115852. The authors benefitted from discussions with, L. Pintschovius, I.I. Mazin, T. Yildirim, R.

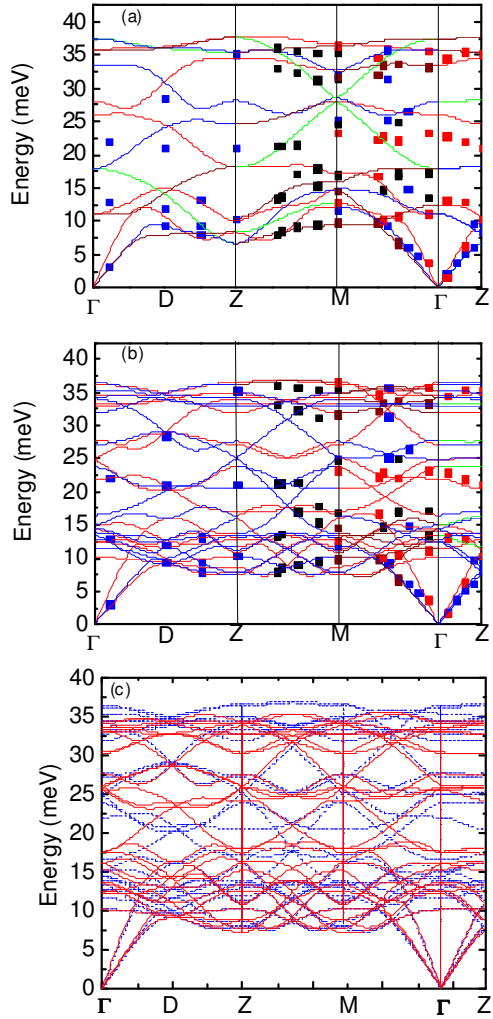

Figure 1. Measured and calculated phonon frequencies. Lines represent phonon dispersions calculated in GGA, whereas the data points represent experimental phonon energies for BaFe$_{1.8}$Co$_{0.2}$As$_2$. The calculations ignored Co-doping since no significant doping dependence has been found in the experiments. (Fig. 4) Experimental and theoretical curves are color coded to reflect different symmetries of the eigenvectors. The experimental points can correspond only to the calculated curves of the same color. In black-and-white the correspondence is for both a and b: solid line - rhomb, dashed line - square, dotted line - none, dashed-dotted – triangle. Black circle indicates that the symmetry of the observed phonons could not be determined.
(a) nonmagnetic calculations with the fully optimized structure compared with experiment.
(b) Same experimental results as in (a) compared with the magnetic calculation with the experimental lattice constants of low temperature of BaFe$_2$As$_2$ and the optimized internal parameter.
(c) Comparison of the calculation in (b) (blue dashed lines) with the nonmagnetic calculation with the structure fixed to the one obtained in the magnetic calculation (red solid lines). This is third calculation (see text) called nonmagnetic orthorhombic in Fig. 3. Including magnetism has the main effect of softening the phonons around 25 meV and hardening the ones around 33 meV. Note that there are two inequivalent Γ-M directions but only the one along the magnetic order propagation vector is shown.

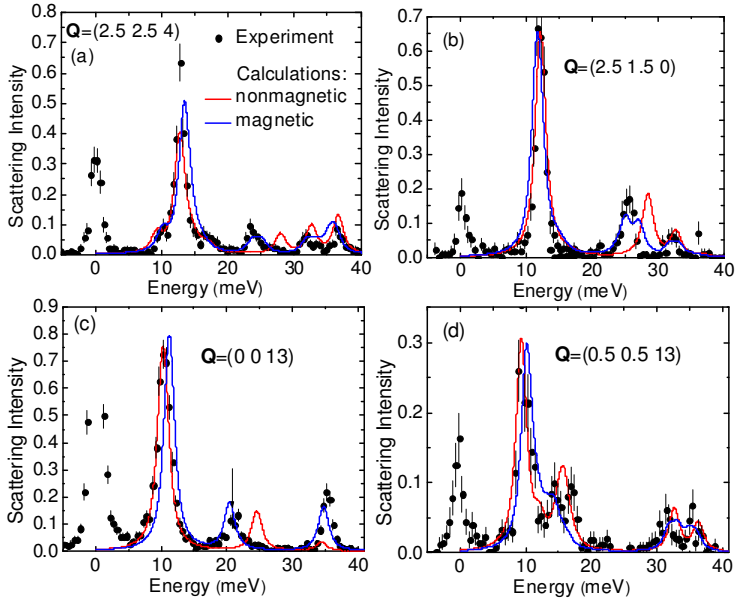

Figure 2. Comparison of measured and calculated inelastic spectra for $BaFe_{1.8}Co_{0.2}As_2$. Arbitrary overall scaling factor has been applied to the calculation. The peak at zero energy comes from elastic scattering. The wavevectors reduced to the first BZ are: (a) (0.5,0.5 0) longitudinal (b) (0.5,0.5) transverse, (c) (0,0,1) longitudinal, (d) (0.5,0.5,1) transverse.

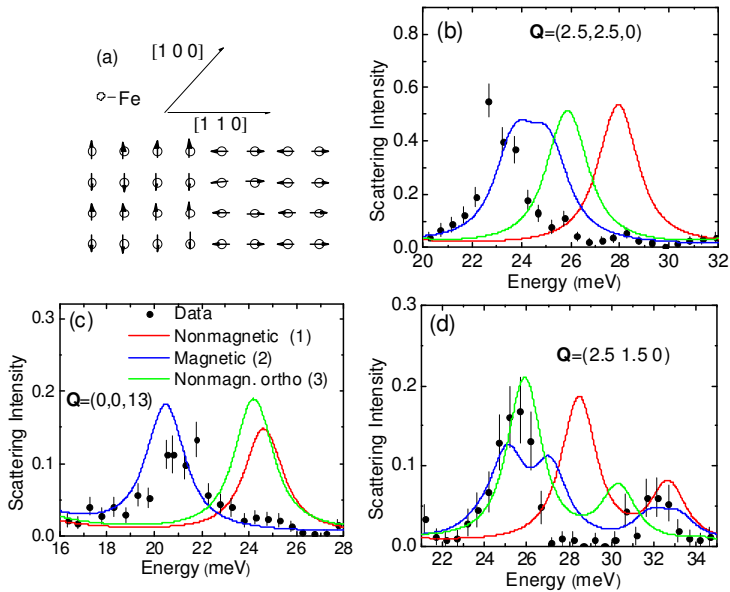

Figure 3 Comparison of the GGA calculations (see text) and the data for phonons expected to be sensitive to magnetism. (a) Schematic of the coupling of phonons to Fe moments in different twin domains. Arrows represent phonon wavevector **q**. Phonons propagating in the [1 1 0] direction interact spins that are oriented differently (0° and 90°) with respect to **q** in the two domains, whereas for the [1 0 0] direction (and for the [0 0 1] direction) the spin orientation relative to **q** is the same for the two domains. (b-d) phonon data (circles) superimposed with calculation results. Red line represents the nonmagnetic calculation with optimized structure. Blue line represents the magnetic calculation with optimized Fe moment with unit cell parameters constrained to the experimental ones for the low temperature phase and relaxed As position. The green line represents the nonmagnetic calculation (zero Fe moment) with the structure fixed to the one obtained from the magnetic calculation. (see text for details)

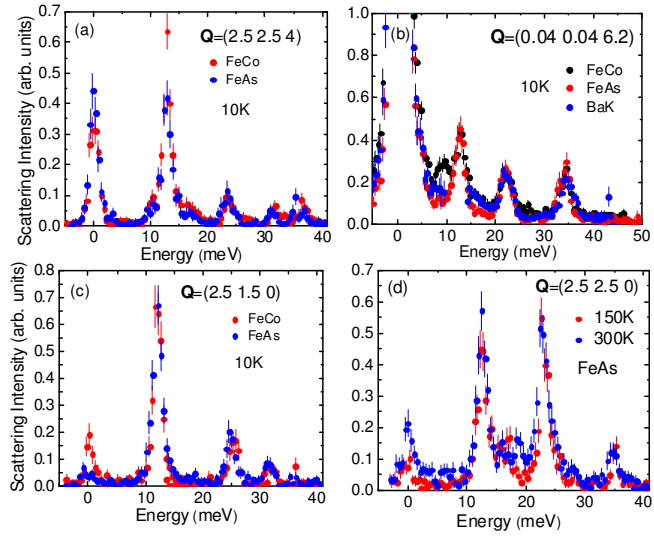

Figure 4 Composition and temperature dependence of the phonon spectra. Except for a slight softening of some a-b plane-polarized phonons in the undoped sample compared to the doped ones, the composition and temperature dependence is generally weak. Small changes such as those near 18meV as well as the feature at 9meV in BaK are typically irreproducible.